\documentclass[%
prd,%
twocolumn%
,secnumarabic%
,amssymb, amsmath,nobibnotes, aps]{revtex4}
\usepackage{epsfig}%
\usepackage{graphicx}%
\usepackage{amsmath}
\usepackage{hyperref}
\usepackage{subfigure}
\expandafter\ifx\csname package@font\endcsname\relax\else
 \expandafter\expandafter
 \expandafter\usepackage
 \expandafter\expandafter
 \expandafter{\csname package@font\endcsname}%
\fi
\begin{document}

\title{Red and blue tilted tensor spectrum from Gibbons-Hawking temperature }

\author{ Subhendra Mohanty $^a$ and Akhilesh Nautiyal $^b$}
\affiliation{$^a$Physical Research Laboratory, Ahmedabad 380009, India.}
\affiliation{$^b$ Institute of Mathematical Sciences,Taramani,
Chennai 600113, India.}
\def\be{\begin{equation}}
\def\ee{\end{equation}}
\def\al{\alpha}
\def\bea{\begin{eqnarray}}
\def\eea{\end{eqnarray}}

\begin{abstract}

The scale invariant scalar and tensor perturbations, 
which are predicted from inflation, are  eigenmodes in the conformal coordinates. The 'out'  observer in the de Sitter space observes a thermal spectrum with a Gibbons-Hawking temperature $H/2\pi$ of these 'Bunch-Davies' particles. The tensor power spectrum observed in  experiments can have an imprint of the Gibbons-Hawking thermal distribution due to the mode mixing between 'in' state conformal coordinates and the coordinate frame of the observer.  We find that the the Bunch-Davies modes appear as thermal modes to the asymptotic Minkowski observer in the future and  the power spectrum of the gravitational waves is blue-tilted with a spectral 
index $n_T \sim 1$ even in the standard slow-roll inflation.  
On the other hand if the coordinate frame of the observer is taken to be static coordinates, the tensor spectrum is red-tilted with $n_T\sim -1$.  A likelihood analysis shows and find  the best fit values  of the slow-roll parameters for both cases. 
We find that the blue-tilted tensor gives a better fit and reconciles the PLANCK upper bound on the tensor-to-scalar 
ratio, $r <0.11$ with BICEP2 measurement of $r=0.2$. This supports the idea of  particle production due to the mode mixing 
between the initial Bunch-Davies vacuum modes and the asymptotic Minkowski vacuum of the post-inflation universe.

\end{abstract}
\maketitle

\section{Introduction}

The prediction of a scale invariant scalar and tensor perturbations \cite{scalar, grw}  from inflation \cite{Inflation} rest on the assumption of a  Bunch-Davies initial state in conformal coordinates of de Sitter space \cite{Mukhanov:2007zz}. An observer in  a different coordinate system, for instance an inertial observer in the static coordinates,  will  see the same perturbations as a 
thermal distribution  with a Gibbons-Hawking temperature $T=H/2\pi$  \cite{Gibbons:1977mu}  due to mode-mixing between the Bunch-Davies modes and eigenmodes of the static coordinates  \cite{Lapedes:1977ip,Mishima:1987uj,Brandenberger:1982xi,Mottola:1984ar,Allen:1987tz,Spradlin:2001pw,Greene,Agullo:2008ka} . Another way by which the scale invariant perturbations produced during inflation can appear as a thermal distribution is when one considers the mode mixing due to the change in observer between the conformal observer during inflation and the asymptotic Minkowski observer in future which measure the perturbations \cite{Polyakov:2012uc,Anderson:2013ila,Singh:2013dia}.  Variations of the standard Bunch-Davies state can be of 
phenomenological interest as a way of reconciling the large value of the tensor-to-scalar ratio implied by the B-mode 
polarization measurement by the BICEP2 collaboration with the lower upper bound established  by PLANCK from the temperature anisotropy \cite{Ashoorioon:2014nta}.

The  BICEP2 collaboration \cite{Ade:2014xna} reported a tensor-to-scalar ratio  $r={0.2}^{+0.07}_{-0.05}$  by the measurement of 
the B-mode polarization \cite{Seljak:1996gy}, which is in apparent contradiction with the upper bound $r<0.11$ (at $95\%$ CL) placed by PLANCK  \cite{Ade:2013uln} from the measurement of the  TT spectrum. There is no direct contradiction between these two measurements  as BICEP2 is most
sensitive at $l\sim 150$ corresponding to a hub of $k=0.01 Mpc^{-1}$ while the PLANCK 2013 measurement uses the hub $k=0.002  Mpc^{-1}$ which corresponds to $l\sim 30$. However explaining the two measurements in a model of inflation would require  (1) a blue tilted tensor spectrum with spectral index $n_T \sim 1 $ \cite{Gerbino:2014eqa,Smith:2014kka,Wang:2014kqa} or (2) a running of the scalar 
spectrum $d \ln n_s/d \ln k = -0.02 $ \cite{Ade:2013uln}. Either of the possible ways to explain the PLANCK-2013 and BICEP2 data 
simultaneously would require going beyond the single field inflation with Bunch-Davies initial state. 
Subsequently the dust popularization measurement reported by PLANCK-2014 \cite{Adam:2014bub} has diminished  the statistical 
significance of the BICEP2 measurement but not ruled it out \cite{Mortonson:2014bja}. There is a possibility that the 
measurement of the B-mode polarization in other experiments (like KECK, SPTpol etc ) may result in a value of the tensor-to-scalar ratio which still calls for a 
non-standard interpretation of the inflationary power spectrum to evade the standard consistency relation $n_T= r /8$ of the 
 standard single field inflation. 

In this paper we show that if we assume a mode mixing between the Bunch-Davies initial vacuum and the post-inflation final vacuum and the Bogoliubov coefficients $\alpha$ and $\beta$   of the mode-mixing is of the thermal form $|\beta|^2=\frac{1}{e^{\beta\omega}-1}$ with the Gibbons-Hawking temperature $T=\beta^{-1}=H/2\pi$, then the spectral index of tensor modes will be blue-tilted 
with  $n_T=1-2\epsilon $.  
On the other hand if we assume that the 'out' observer is the one with the static 
coordinates, then 
the Bogoliubov coefficients again give the same  thermal distribution with 
identical $ |\beta|^2$, however, the spectral index in 
this case is red-tilted $n_T=-1-2\epsilon$. The difference in the spectral tilt between the two cases is due to the fact that 
when we transform the initial state from the conformal to static coordinates we have $\alpha \beta^* <0$ while the 
transformation between the conformal coordinates and the asymptotic Minkowski coordinates of the late time observer gives 
$ \alpha \beta^*>0$. The difference in sign of $\alpha \beta^*$ with the same $|\beta|^2$ results in a different spectral 
tilt. In order to avoid the successful prediction of the scale invariant scalar power spectrum, we will 
assume that the slow roll parameter $\eta$ of the scalar potential is negative so that the scalar modes are tachyonic and 
the Hawking radiation of scalar modes is suppressed \cite{Epstein:2014jaa}.

We do a likelihood analysis for the values of the tensor-to-scalar ratio for the case of red and blue tilted spectra and 
determine the slow roll parameters of the model which would be reconcile the B-mode and TT anisotropy data. 
We conclude that mode mixing between the Bunch-Davies vacuum and the
vacuum state of the observer, may resolve the tension between the PLANCK-2013 bound and BICEP2 measurement and the accurate 
experimental measurement of the spectral index can determine the nature of the initial state of the inflation generated perturbations.

\section{Bogoliubov transformation of Bunch-Davies vacuum}

We can express the tensor perturbations $h({\bf x},t)$ as a quantum field in terms of the mode functions
$\phi_{in\, k}$ (which
satisfies the minimally coupled Klein-Gordon equation) as
\be
h({\bf x},t)=\frac{\sqrt{2}}{M_p}\int
[dk] \,(  a_{k} \, \phi_{in\, k}+ a_{k}^\dagger \, \phi_{in\, k}^\star),
\label{bdexprs}
\ee
where $a_{k}^\dagger$ ($a_{k}$) are the creation (annihilation)
operators of the 'particles' in the conformal vacuum, also called the Bunch-Davies vacuum, which we will denote 
by $|0_{in}\rangle$ and which is defined by $a_{k}|0_{in}\rangle=0$.  Eqn.~(\ref{bdexprs}) is written in terms of
the spherical polar coordinates.

The Bunch-Davies vacuum is defined in conformal
coordinates $(\eta, \rho, \theta, \phi)$ with the line element
\bea ds^2&=& \frac{1}{H^2 \eta^2} \left( d\eta^2 - d\rho^2-\rho^2 d\Omega^2 \right)\,
\nonumber\\
\eta &\in& (-\infty,0),\,\,\,\,\rho\in (0,\infty).
\label{conformal}
\eea

The mode functions $\phi_{in\, k}$  are solutions  of the Klein-Gordon equation in the conformal coordinates,
\be
\frac{\partial^2 \phi_{in\, k}}{\partial
\eta^2}-\frac{2}{\eta}\frac{\partial \phi_{in\, k} }{\partial \eta} -
{\nabla_\rho}^2 \phi_{in\, k}=0.
\label{kgbd}
\ee
This equation has the exact solution
\be
\phi_{in\, k}(\eta, \rho, \theta, \phi)= \frac{iH}{\sqrt{2k^3}}\,e^{-ik\eta}\,
\left(1+ik \eta \right)\, j_l(k\eta)\, \frac{Y_{l,m}(\theta,
\phi)}{\sqrt{4 \pi}}.
\label{bdsol}
\ee
The zero-point fluctuations during inflation are assumed to be eigenmodes of the 
KG equations in conformal coordinates, as the mode functions (\ref{bdsol}) in the 
high $k$ limit have the same form $\phi_{in\, k}\sim \frac{1}{\sqrt{2k}}\, e^{-i(k \eta - {\bf k\cdot x })}$ as positive frequency modes in Minkowski space.

The scalar
field is quantized in terms of the creation and annihilation
operators of the $\phi_{out\, k}$ quantum modes which are the elementary
excitations in the different coordinate system dependent on the observer,
\be
h({\bf x},t)=
\int [d\omega] \,( b_{\bf \omega} \, \phi_{out\, \omega}+ 
b_{\bf \omega}^\dagger \, \phi_{out\, \omega}^\star).
\ee
Here $b_{\bf \omega}^\dagger$ and $b_{\bf \omega}$ are the creation and annihilation
operators acting on a different vacuum $|o_{out}\rangle$.
The two sets of modes $\phi_{in\, k}$ and $\phi_{out\, \omega}$  can be
linearly related in terms of Bogoliubov coefficients $(\alpha_{\omega k},\, \beta_{\omega k})$
as
\bea
\phi_{out\, \omega}&\equiv&\int[dk]\left(\alpha_{\omega k}\phi_{in\, k}+
\beta_{\omega k}\phi_{in\, k}^\star\right)\nonumber\\
 \phi_{in\, k} &\equiv&\int[d\omega]\left(\alpha_{\omega k}^\star \,
\phi_{out\, \omega} - \beta_{\omega k}\, \phi_{out\, \omega}^\star\right)
\label{bogodef}\eea
The relation (\ref{bdexprs}) in terms of cartesian coordinates $(\eta,x,y,z)$ can
be written as
\bea
h({\bf x},\eta)&=&\frac{\sqrt{2}}{ M_p}\int \frac{d^3k}{(2\pi)^{3/2}}h({\bf k},\eta)e^{i {\bf k}\cdot {\bf x}}\nonumber\\
&=&\frac{\sqrt{2}}{ M_p}\int \frac{d^3k}{(2\pi)^{3/2}}
\left(a_{{\bf k}}\phi_{in\, k}+a_{{-\bf k}}^\dagger \phi_{in\, k}^\star\right)
e^{i {\bf k}\cdot {\bf x}}.\nonumber\\
\label{hx}
\eea
The power spectrum of tensor perturbation is given in terms of two-point 
correlation function of the field $h({\bf x},\eta)$, which in the out-vacuum 
$|0_{out}\rangle$ can be obtained as
\bea
&&\langle 0_{out}|h({\bf x},\eta)h({\bf y},\eta)|0_{out}\rangle\nonumber\\
&&=
\frac{2}{{M_P}^2}\int \frac{d^3kd^3k^\prime}{(2\pi)^{3}}
\langle 0_{out}|h({\bf k},\eta)h({\bf k^\prime},\eta)|0_{out}\rangle
e^{i\left( {\bf k}\cdot {\bf x}+{\bf k^\prime}\cdot {\bf y}\right)}.\nonumber\\
\eea

To compute the  two-point correlation function 
$\langle 0_{out}|h({\bf k},\eta)h({\bf k^\prime},\eta)|0_{out}\rangle$
we will again use the spherical coordinates $(\eta,r,\theta,\phi)$.
 Now we can express the creation and annihilation operators of the in-vacuum
$a_{{k}},\, a_{ k}^\dagger$ in terms of the creation and annihilation
operators of the out-vacuum $b_{\omega},\, b_{\omega}^\dagger$ using
Eqn.~(\ref{bogodef}) as
\bea
a_{{ k}}&=&\int[d\omega]\left(
\alpha_{\omega k}\, b_{{ \omega}}+\beta^\star_{\omega k}\,
 b_{{\omega}}^\dagger\right)\nonumber\\
a_k^\dagger&=&\int [d\omega]\left(\beta_{\omega k}\,  b_{{\omega}} +
\alpha^\star_{\omega k}\,  b_{{\omega}}^\dagger\right).
\label{bt1}
\eea
 So we obtain
\bea
&&\langle 0_{out}|h({\bf k},\eta)h({\bf k^\prime},\eta)|0_{out}\rangle\nonumber\\
&&=\delta\left({ k}-{ k^\prime}\right)
\int [d\omega][d\omega^\prime]\left[\left(\alpha_{\omega k}
\alpha^\star_{\omega^\prime k}+\beta_{\omega k}\beta^\star_{\omega k}\right)
|\phi_{in\, k}|^2\right.\nonumber\\&&\left.+
\left(\alpha_{\omega k} \beta^\star_{\omega^\prime k} (\phi_{in\, k})^2\right)
+\left(\alpha^\star_{\omega k} \beta_{\omega^\prime k} (\phi^\star_{in\, k})^2\right)
\right].\nonumber\\
\label{hhk}
\eea
Now for the choices of out-vacua we consider in the next sections, 
$\alpha_{\omega,k}$ and $\beta_{\omega,k}$ are diagonal in $\omega$ and the 
frequency  $\omega$ of the out-vacuum corresponds to $\frac{k}{a}$ so the integrals 
in above expressions can be done by using $\delta(\omega-\frac{k}{a})$.
Now using the temporal part of $\phi_{in\, k}$ given be Eqn.~(\ref{bdsol})
\be
\phi_{in\, k}=\frac{iH}{\sqrt{2 k^3}}\left(1+ik\eta\right)e^{-i k\eta},
\label{ukin}
\ee
 we get for the super-horizon
($k \eta \ll 1$) perturbations
\bea
&&\langle 0_{out}|h({\bf k},\eta)h({\bf k^\prime},\eta)|0_{out}\rangle\nonumber\\
&=&\delta\left({ k}-{ k^\prime}\right)\frac{H^2}{k^3 M_p^2}
\left[|\alpha_{\omega k}|^2+|\beta_{\omega k}|^2+
2Re\left(\alpha_{\omega k} \beta_{\omega k}^\star \right)\right].\nonumber \\
\label{hh2}
\eea
Now the tensor power spectrum is defined as
\be
4\times \frac{k^3}{2\pi^2}\langle 0_{out}|h({\bf k},\eta)h({\bf k^\prime},\eta)|0_{out}
\rangle=\delta\left({ k}-{ k^\prime}\right)P_T(k).
\ee
So
\bea
P_T&=&\!\frac{8}{M_P^2}\! \left(\frac{H}{2 \pi}\right)^{2}\!  \left(\frac{k}{a H}\right)^{-2\epsilon} \!\left[|\alpha_{\omega k}|^2+|\beta_{\omega k}|^2\right.\nonumber\\
&+&\left. 2Re\left(\alpha_{\omega k} \beta_{\omega k}^\star \right)\right].
\label{Power}
\eea

\section{ Tensor Power spectrum measured for the Static Observer}

The coordinate system which describes the
 the static observer in de Sitter space with coordinate
$(t,r,\theta,\phi)$ and the metric given by
\bea ds^2&=& (1-r^2 H^2) dt^2 -\frac{1}{(1-r^2 H^2)} dr^2-r^2
d\Omega^2 \,
\nonumber\\
t&\in& (-\infty,\infty),\,\,\,\,r \in (0,H^{-1})\,.
\label{staticco}
\eea
The time evolution of the quantum state with respect  to an
observer located at $r=0$ is determined by a Hamiltonian operator
defined by the time-like Killing vector $\partial_t$. The static
coordinate system has a coordinate singularity at $r=H^{-1}$ which
is the event horizon for the observer at $r=0$.

The two coordinate systems overlap in the region $\eta \in
(-\infty,0)$ and can be related as
\bea
\eta &=&-\frac{1}{H\left(1-H^2r^2\right)^{\frac{1}{2}}}e^{-H t},\nonumber\\
\rho &=& - r \eta .
\label{transbdso}
\eea  
We consider first the two-dimensional spacetime to obtain the Bogoliubov coefficients for an static observer. We
will also set $H=1$ to make the notation simple and  at the end of the calculation we will restore $H$ by substituting 
$\left(r, t\right)\rightarrow \left(Hr, Ht\right)$ and $k\rightarrow \frac{k}{H}$.  The metric (\ref{staticco}) 
for two-dimensional static coordinates becomes 
\bea ds^2&=& (1-r^2 H^2) dt^2 -\frac{1}{(1-r^2 H^2)} dr^2
\nonumber\\
t&\in& (-\infty,\infty),\,\,\,\,r \in (-H^{-1},H^{-1})\,.
\label{static2d}
\eea
The solution of Klein-Gordon eqn. $\square \phi=0$ in this coordinate system is given as
\be
\phi_{out\, \omega}(t, r)=\frac{1}{\sqrt{2\omega}}e^{-i\omega t}{\left[\frac{1+r}{1-r}\right]}^{\frac{i\omega}{2}},
\label{phiout2d}
\ee
 and the solution of the KG eqn.~(\ref{kgbd}) in sub-Hubble limit in two-dimension is given as 
\be
\phi_{in\, k}(\eta,\rho)=\frac{1}{\sqrt{2 k}}e^{-ik\left(\eta-\rho\right)}.
\label{phiin2d}
\ee
Now the Bogoliubov coefficients defined in Eqn.~(\ref{bogodef}) can be obtained by Klein-Gordon inner product \cite{BD}
\bea
\alpha_{\omega,k}&=&\langle{\phi_{out\, \omega}\phi_{in\, k}}\rangle\nonumber\\
\beta_{\omega,k}&=&-\langle{\phi_{out\, \omega}\phi_{in\, k}^\star}\rangle.
\label{bogoinp}
\eea
Here the Klein-Gordon inner product is defined as
\be
\langle\phi_{out\, \omega}\phi_{in\, k}\rangle=-i\int_t dr \sqrt{-g}g^{0\nu}
\phi_{out\, \omega}\overleftrightarrow{\partial_\nu}\phi_{in\, k}^\star.
\ee
Now using the metric (\ref{static2d}) and integrating over the constant time hypersurface we can obtain $\alpha_{\omega,k}$ as
\be
\alpha_{\omega,k}=-i\int_{-1}^1 \frac{dr}{1-r^2}
\left(\phi_{out\, \omega}\overleftrightarrow{\partial_0} \phi_{in\, k}^\star\right)
\label{alphainp}
\ee
The Bunch-Davies mode $\phi_{in\, k}$ can be expressed in terms of the static coordinates $r,\,  t$ using the transformations
(\ref{transbdso}) as
\be
\phi_{in\, k}=\frac{1}{\sqrt{2k}}e^{ik\left[\frac{1+r}{1-r}\right]^{\frac12}e^{-t}}
\label{phiinstat}
\ee
Now using the modes (\ref{phiinstat}) and (\ref{phiout2d}) and evaluating the integral (\ref{alphainp}) at $t=0$ 
hypersurface we get 
\bea
\alpha_{\omega,k}&=&\frac{1}{2\sqrt{\omega k}}\int^1_{-1}\frac{dr}{1-r^2}\left[k\left(\frac{1+r}{1-r}\right)^{\frac{1+i\omega}{2}}
e^{-ik\left(\frac{1+r}{1-r}\right)^{1/2}}\right.\nonumber\\
	&+&\left.\omega\left(\frac{1+r}{1-r}\right)^{\frac{i\omega}{2}}
e^{-ik\left(\frac{1+r}{1-r}\right)^{1/2}}\right]
\eea
Now changing variable to  $z=\left(\frac{1+r}{1-r}\right)^{1/2}$ the above integral becomes
\be
\alpha_{\omega,k}=\frac{1}{2\sqrt{\omega k}}\int^\infty_{0} dz\left(k z^{i\omega}+\omega z^{i\omega-1}\right)e^{-ikz}.
\ee
This integral can be solved using the $\Gamma$ functions and finally we get
\be
\alpha_{\omega,k}=\sqrt{\left(\frac{\omega}{k}\right)}k^{-i\omega}e^{\frac{\pi\omega}{2}}\Gamma{(i\omega)}.
\label{alphastat}
\ee
Similarly we obtain the another coefficient $\beta_{\omega,k}$ using the inner product (\ref{bogoinp}) as
\be
\beta_{\omega,k}=-\sqrt{\left(\frac{\omega}{k}\right)}k^{-i\omega}e^{-\frac{\pi\omega}{2}}\Gamma{(i\omega)}
\label{betastat}
\ee 

 Now in four-dimension to solve the Klein-Gordon equation for the 
 modes of static observer $\phi_{out\, \omega}$ 
we separate wave function in the
$(t,r,\theta,\phi)$ coordinates
\be \phi_{out\, \omega}=\frac{ f(r)}{r} Y_{l,m}(\theta, \phi)e^{-i \omega t}.
\ee 

  The equation for radial wave function
can be written in a simple form
\be
\frac{d^2}{dr_*^2} f(r)+(1-r^2)\left(\frac{l(l+1)}{r^2}-2\right)f(r)
+k^2 f(r)=0, \label{radial}
\ee
in terms of the tortoise coordinates \be r_* \equiv \int
\frac{dr}{(1-r^2)} =\frac{1}{2}\ln\left(\frac{1+r}{1-r}\right).
\ee
The radial equation can be solved exactly in terms of
Hypergeometric functions that can be written in terms of Legendre functions of second kind and the solution is
\be
\phi_{out\, \omega}(t,r,\theta,\phi)=
\frac{1}{\sqrt{2\omega}}e^{-i\omega t}Q^{i\omega}_l\left[\frac{1}{r}\right]
Y_{lm}(\theta,\phi).
\label{static}
\ee

The sub-Hubble limit
of the Bunch-Davies mode functions (\ref{bdsol}) is given as
 \be
\phi_{in\, k}=-\frac{1}{\sqrt{2k}}\eta e^{-ik\eta}j_l(k\eta)Y_{lm}(\theta,\phi).
\ee
It can be shown (see \cite{Mishima:1987uj}) that the Bogoliubov coefficients for
$l=0$ in four-dimensions are same as for two-dimensions except normalization factors.
So we can write the Bogoliubov transformations for $l=0$ case in four-dimensions as
\bea
\alpha_{\omega,k}&=&N k^{-i\omega} e^{\frac{\pi\omega}{2}}\Gamma(i\omega),\nonumber\\
\beta_{\omega,k}&=&-N k^{-i\omega} e^{-\frac{\pi\omega}{2}}\Gamma(i\omega). \label{bogo4d}
\eea
Here $N$ is normalization constant. Putting back $k=k/H$ and $\omega=\omega/H$, using the identity
\be
|\Gamma{(i\omega)}|^2=\frac{\pi}{\omega\sinh(\pi\omega)}
\ee
and  the normalization condition for Bogoliubov coefficients 
\be
\int [dk]\left(\alpha_{\omega k}\alpha^\star_{\omega^\prime k}-
\beta_{\omega k}\beta^\star_{\omega^\prime k}\right)=\delta(\omega-\omega^\prime),
\label{normbogo}
\ee
we obtain the expressions for $\alpha_{\omega k}$ and $\beta_{\omega k}$ from Eqn.~ (\ref{bogo4d}) as
\bea
|\alpha_{\omega k}|^2&=&\frac{e^{\beta\omega }}{e^{\beta\omega}-1}\nonumber\\
|\beta_{\omega k}|^2&=&\frac{1}{e^{\beta\omega}-1}\nonumber\\
\alpha_{\omega k}\beta_{\omega k}^\star&=&-
\frac{e^{\frac{\beta\omega}{2}}}{e^{\beta\omega}-1},
\label{finalbogostatic}
\eea
where
\be
\beta=\frac{2\pi}{H}.
\ee

Using  $\alpha_{\omega k}$ and $\beta_{\omega k}$ given by
(\ref{finalbogostatic}) in two-point function (\ref{hh2}) the tensor power spectrum
can be expressed as
\bea
P_T&=&\frac{8}{M_P^2}\left(\frac{H}{2 \pi}\right)^{2} \left(\frac{k}{a H}\right)^{-2\epsilon} 
\left[ \frac{(e^{\frac{\pi k}{a H}}+1)^2}{(e^{\frac{2\pi k}{a H}}-1)}\right]\nonumber\\
&\simeq&\frac{8}{M_P^2}\left(\frac{H}{2 \pi}\right)^{2} \left(\frac{k}{a H}\right)^{-2\epsilon} \left[\frac{2}{\pi}\left(\frac{aH}{k}\right)\right] \, {\rm for} \, k\ll aH,\nonumber\\
\label{Power_red}
\eea
which is a red-tilted spectrum for the tensor modes with spectral index $n_T=-1-2 \epsilon$.

\section{ Particle production in late Universe}
The post inflation universe at the time when all the modes are sufficiently sub-horizon, the modes can be considered as  plane waves in  Minkowski space,
\be
\phi_{out\, \omega}=A\,e^{-i \omega t},
\ee
while the Bunch-Davies modes (\ref{ukin}) in the sub-horizon limit is
\bea
\phi_{in\, k}&=&\frac{1}{a} \frac{1}{\sqrt{2k}}\,e^{-i k \eta}\nonumber\\
&=& B \, e^{-t}\,e^{ik e^{-t}}.
\eea
Using  Bogoliubov transformations (\ref{bogodef}) we get
\bea
\phi_{in\, k}&=&A\int_0^\infty d\omega\left(\alpha^\star_{\omega,k}e^{-i\omega t}
-\beta_{\omega,k}e^{i\omega t}\right)\nonumber\\
&=&A\int_{-\infty}^{\infty}d\omega f(\omega,k)e^{-i\omega t}.
\eea
So
\be
\alpha^\star_{\omega,k}=f(\omega,k),\, \, \, \beta_{\omega,k}=-f(-\omega,k).
\ee
So the Bogoliubov coefficients for these 'in' and 'out' states can be
obtained by doing inverse Fourier Transforms as
\bea
\alpha_{\omega,k}^*&=&\frac{1}{A} \int_{-\infty}^{\infty} \phi_{in}(k) e^{i \omega t} dt\nonumber\\
&=& \frac{B}{A}  \int_{-\infty}^{\infty} e^{-t}\,e^{ik e^{-t}} e^{i \omega t} dt.
\eea
By substituting $e^{-t}=z$ the integral reduces to
\bea
\alpha_{\omega,k}^*&=&\frac{B}{A} \int_{0}^{\infty} z^{-i \omega}e^{ik z} dz\nonumber\\
&=&\frac{B}{A} \, \frac{\omega}{k} \,k^{-i \omega} e^{\pi \omega/2} \Gamma(-i\omega).
\eea
Similarly
\bea
\beta_{\omega,k}&=&-\frac{1}{A} \int_{-\infty}^{\infty} \phi_{in}(k) e^{-i \omega t} dt\nonumber\\
 &=&\, \frac{\omega}{k} \,k^{-i \omega} e^{\pi \omega/2} \Gamma(-i\omega)
\eea
 
 Normalizing $\alpha_{\omega k}$ and $\beta_{\omega k}$ we obtain
\bea
|\alpha_{\omega k}|^2&=&\frac{e^{\beta\omega }}{e^{\beta\omega}-1}\\
|\beta_{\omega k}|^2&=&\frac{1}{e^{\beta\omega}-1}\\
\alpha_{\omega k}\beta_{\omega k}^\star&=&
\frac{e^{\frac{\beta\omega}{2}}}{e^{\beta\omega}-1},
\label{finalbogominko}
\eea
where again
$\beta=\frac{2\pi}{H}$, which is the Hawking-Gibbon temperature of the de Sitter space..
Using  $\alpha_{\omega k}$ and $\beta_{\omega k}$ given by
(\ref{finalbogominko}) in two-point function (\ref{hh2}) the tensor power spectrum
can be expressed as
\bea
P_T&=&\frac{8}{M_P^2}\left(\frac{H}{2 \pi}\right)^{2} \left(\frac{k}{a H}\right)^{-2\epsilon} \left[ \frac{(e^{\frac{\pi k}{a H}}-1)^2}{(e^{\frac{2\pi k}{a H}}-1)}\right]\nonumber\\
&\simeq&\frac{8}{M_P^2}\left(\frac{H}{2 \pi}\right)^{2} \left(\frac{k}{a H}\right)^{-2\epsilon} \left[\frac{\pi}{2}\left(\frac{k}{aH}\right)\right] \, {\rm for} \, k\ll aH,\nonumber\\
\label{Power_blue}
\eea
which is a blue-tilted spectrum for the tensor modes with $n_T=1-2 \epsilon$.

\section{Best fit values of r for red and blue tilted tensor spectrum}

\begin{figure}[h!]
\centering
\includegraphics[width=6cm,height=7cm,angle=-90]{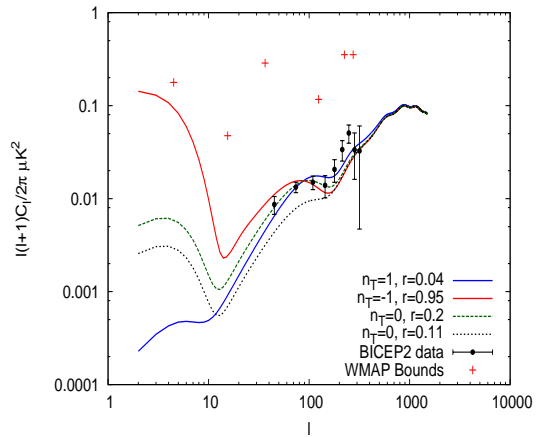}
\caption{$B$-modes from modified as well as standard power spectrum
with BICEP2 data and WMAP bounds}
\label{bicephwbb}
\end{figure}

We use the  power spectra (\ref{Power_red}) and (\ref{Power_blue}) to compute the angular power spectra for $B$-mode polarization using CAMB \cite{Lewis:1999bs}.
We have taken the best-fit values of the parameters
($\Omega_bh^2$, $H_0$, $\Omega_ch^2$, $\tau$, $A_s$ and $n_s$)  given by
PLANCK \cite{Ade:2013zuv} for the base $\Lambda$CDM model. The scalar amplitude $A_s$ and the scalar
spectral index are taken at the pivot scale $k=0.05$Mpc\textsuperscript{-1}.
We have also taken into account  the lensed $B$-modes generated from $E$-modes.
$B$-modes due to the modified and the standard power spectra, along with
BICEP2 data and WMAP bounds are shown in Fig.~\ref{bicephwbb}. The value of
tensor-to-scalar ratio $r$ is taken at the pivot scale
$k=0.002$Mpc\textsuperscript{-1}. We see that with $n_T\simeq 1$
the BICEP2 data is in agreement with a low value
of $r=0.04$ consistent with the  PLANCK's upper bound. In Fig.~\ref{bicephwbb} we
have also shown the standard power spectra (taking $n_T\simeq 0$)
with $r=0.2$ and $r=0.11$ for comparison. We see that in the standard power spectrum,
the PLANCK upper bound $r=0.11$ is not in agreement with the BICEP2 data.

\begin{figure}[h!]
\centering
\includegraphics[width=6cm,height=7cm,angle=-90]{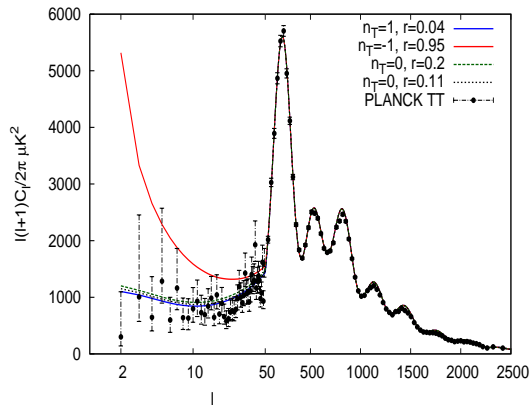}
\caption{$TT$ power spectrum with modified as well as standard tensor power spectrum
 with PLANCK data}
\label{planckhwtt}
\end{figure}

The tensor power also contributes to the temperature anisotropy. In Fig.~\ref{planckhwtt}
we show the $TT$ anisotropy with the modified and the standard power spectra.
As expected the $n_T=1$, $r=0.04$ power spectrum gives the lowest contribution to
temperature anisotropy. With the temperature anisotropy the red-tilted power spectrum (\ref{Power_red}) is ruled out.

\begin{figure}[ht]
\centering
\subfigure[]{
\includegraphics[width=4cm,height=5cm,angle=-90]{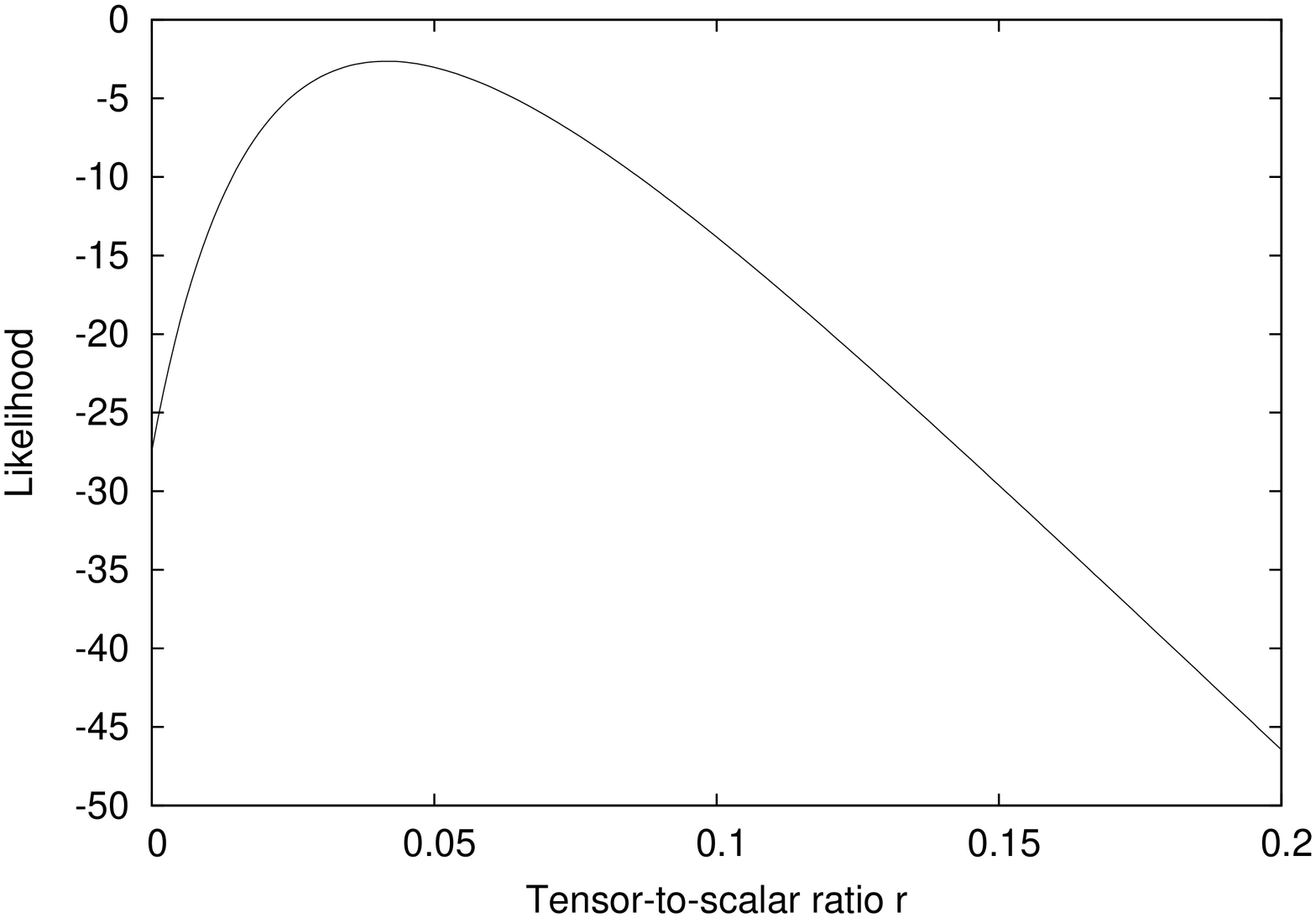}
\label{likelihood_blue}
}
\subfigure[]{
\includegraphics[width=4cm,height=5cm,angle=-90]{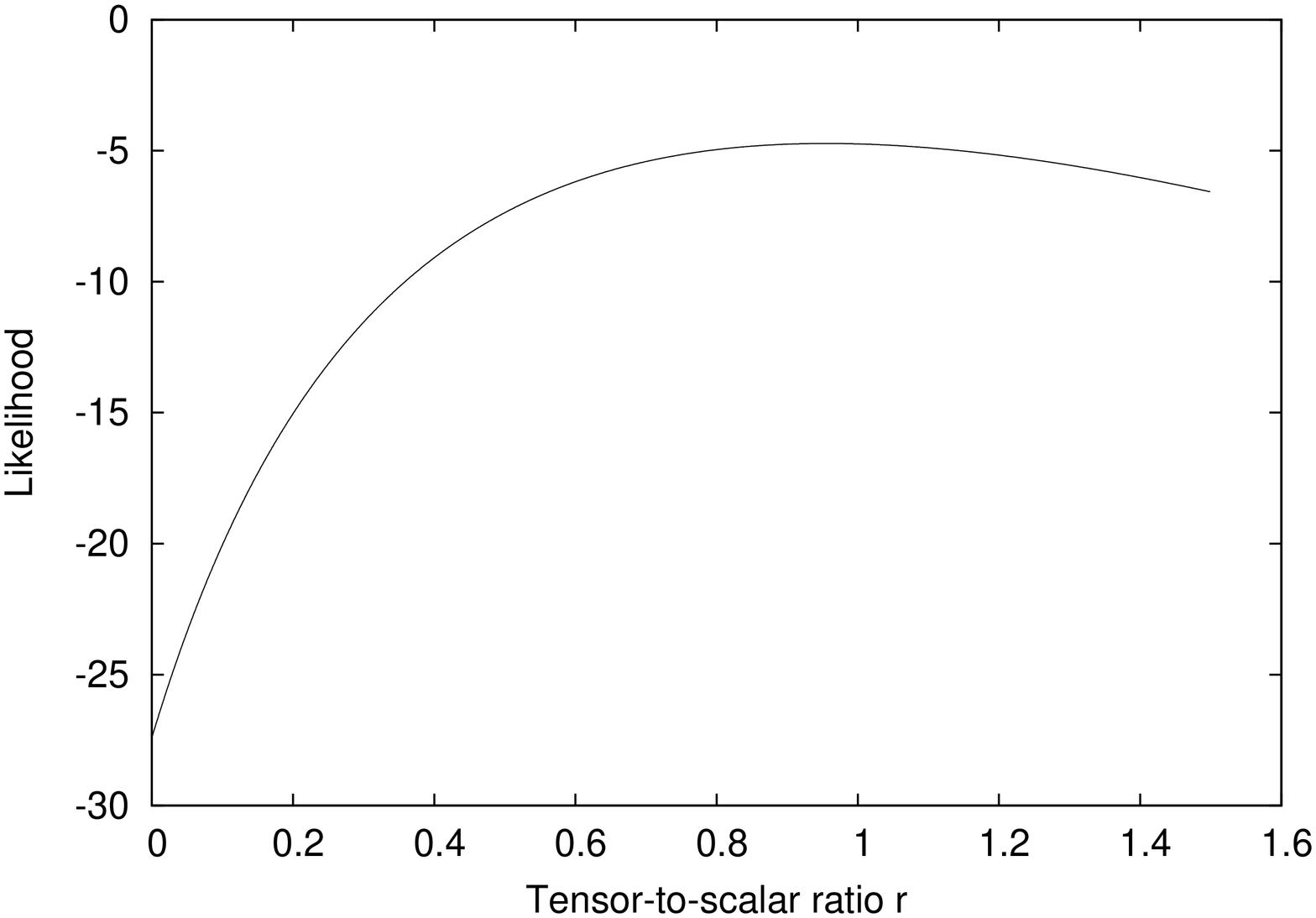}
\label{likelihood_red}
}
\caption{Likelihood for tensor-to-scalar ratio with modified power
spectra (\ref{Power_blue}) (Fig.~{\ref{likelihood_blue}}), (\ref{Power_red}) (Fig.~{\ref{likelihood_red}}). 
The maximum likelihood is at $r=0.042$ for blue-tilted and  $r=0.95$ for red-tilted
power spectrum.}
\label{likelihood}
\end{figure}

In Fig.\ref{likelihood}  we plot the likelihood for the
tensor-to-scalar ratio $r$ using the modified power spectra (\ref{Power_blue}) and (\ref{Power_red}) for the
BICEP2 data. The best fit value of tensor-to-scalar ratio $r$, the 
corresponding maximum likelihood ($\ln {\mathcal L}$) and the slow-roll parameters
for both blue-tilted and red-tilted spectra are given in table \ref{table1}.

 \begin{table*}
\begin{tabular}{|c|c|c|c|c|}
\hline
$n_T$ & $r$ & $\ln {\mathcal L}$ & $\epsilon$ & $\eta$\\
\hline
1 & 0.042 & -2.6390 & 0.002 & -0.014\\
0 & 0.215   & -3.6830 & 0.013 & 0.019\\
-1 & 0.95 & -4.7223 & 0.093 & 0.2607 \\
\hline
\end{tabular}
\caption{Best fit tensor-to-scalar ratio and Maximum Likelihood for blue and 
red-tilted tensor power spectra using BICEP2 data.}
\label{table1}
\end{table*}

As displayed  in Table.~\ref{table1}  the maximum likelihood is at $r=0.042$ for the 
blue-tilted power spectrum and $r=0.95$ for the
red-tilted power spectrum. The value of $\ln {\mathcal L}$ is highest for the 
blue-tilted power spectrum, which shows that the blue-tilted tensor power spectrum
is a better fit to BICEP2 data compared to the red-tilted and scale invariant power 
spectra.

For the maximum likelihood value of  $r_{0.002}=0.042$ from a blue tilted spectrum,
the slow-roll parameter $\epsilon=0.002$. From the scalar spectral index
 $n_s=1-6\epsilon +2\eta=0.9619$ we have $\eta\sim -0.014$.

On the other hand the maximum likelihood value of  $r_{0.002}=0.95$ from a red tilted spectrum,
the slow-roll parameter $\epsilon=0.093$. From the scalar spectral index
 $n_s=1-6\epsilon +2\eta=0.9619$ we have $\eta\sim 0.2607 $.
 
 The accurate determination of the tensor spectrum in future experimental measurements of the B-model  will determine the parameters of the inflation model
 and help in picking out the correct model of inflation.  

%

\section{Conclusion}

The combined data from B-mode measurement by BICEP2 \cite{Ade:2014xna} with the temperature anisotropy measurement from PLANK-2013 \cite{Ade:2013uln} implies that the slow roll inflation consistency relation $n_T\sim r/8$ is violated. It is well known that 
assuming a different initial state compared to the Bunch-Davies one can modify the relation between the 
slow-roll $\epsilon$ parameter derived from the potential and the observed tensor spectral index  $n_T$ \cite{Hui:2001ce,Ashoorioon:2014nta}. 
In this paper we  examine the modification to the tensor spectrum due to mode mixing between a Bunch-Davies 'in' vacuum and 
the 'out' vacuum of the (a) static coordinate observer and (b) the post inflation asymptotic Minkowski observer. 
Both the scenarios result in a Gibbons-Hawking thermal distribution as observed w.r.t the 'out' vacuum.  
The relative phases of the Bogoliubov coefficients are different in the two cases and these lead to quite different predictions 
for the tensor spectral index. The combined BICEP2 and PLANCK-2013 data gives a better fit for a blue-tilted tensor 
spectrum which supports the post-inflation particle production scenario. The Hawking-Gibbons temperature unlike 
temperature of perturbation form a possible pre-inflation radiation era does not go down exponentially during the course of 
inflation so the effect is not diluted after a few e-foldings \cite{Bhattacharya:2005wn, Bhattacharya:2006dm}. 
Measurements of the B-mode in future experiments may give a signature of the Gibbons-Hawking temperature.



\end{document}